RESEARCH ARTICLE  Open Access

# Spatiotemporal clustering, climate periodicity, and social-ecological risk factors for dengue during an outbreak in Machala, Ecuador, in 2010

Anna M Stewart-Ibarra[1*], Ángel G Muñoz[2,3], Sadie J Ryan[1,4,5], Efraín Beltrán Ayala[6,7], Mercy J Borbor-Cordova[8], Julia L Finkelstein[9,10], Raúl Mejía[11], Tania Ordoñez[6], G Cristina Recalde-Coronel[8,11] and Keytia Rivero[11]

## Abstract

**Background:** Dengue fever, a mosquito-borne viral disease, is a rapidly emerging public health problem in Ecuador and throughout the tropics. However, we have a limited understanding of the disease transmission dynamics in these regions. Previous studies in southern coastal Ecuador have demonstrated the potential to develop a dengue early warning system (EWS) that incorporates climate and non-climate information. The objective of this study was to characterize the spatiotemporal dynamics and climatic and social-ecological risk factors associated with the largest dengue epidemic to date in Machala, Ecuador, to inform the development of a dengue EWS.

**Methods:** The following data from Machala were included in analyses: neighborhood-level georeferenced dengue cases, national census data, and entomological surveillance data from 2010; and time series of weekly dengue cases (aggregated to the city-level) and meteorological data from 2003 to 2012. We applied LISA and Moran's I to analyze the spatial distribution of the 2010 dengue cases, and developed multivariate logistic regression models through a multi-model selection process to identify census variables and entomological covariates associated with the presence of dengue at the neighborhood level. Using data aggregated at the city-level, we conducted a time-series (wavelet) analysis of weekly climate and dengue incidence (2003-2012) to identify significant time periods (e.g., annual, biannual) when climate co-varied with dengue, and to describe the climate conditions associated with the 2010 outbreak.

**Results:** We found significant hotspots of dengue transmission near the center of Machala. The best-fit model to predict the presence of dengue included older age and female gender of the head of the household, greater access to piped water in the home, poor housing condition, and less distance to the central hospital. Wavelet analyses revealed that dengue transmission co-varied with rainfall and minimum temperature at annual and biannual cycles, and we found that anomalously high rainfall and temperatures were associated with the 2010 outbreak.

**Conclusions:** Our findings highlight the importance of geospatial information in dengue surveillance and the potential to develop a climate-driven spatiotemporal prediction model to inform disease prevention and control interventions. This study provides an operational methodological framework that can be applied to understand the drivers of local dengue risk.

* Correspondence: stewarta@upstate.edu
[1]Department of Microbiology and Immunology, Center for Global Health and Translational Science, State University of New York Upstate Medical University, 750 East Adams St, Syracuse, NY 13210, USA
Full list of author information is available at the end of the article

## Background

Dengue fever is the most significant mosquito-borne viral disease globally, and has rapidly increased in incidence, geographic distribution, and severity in recent decades [1-3]. The disease is caused by four distinct dengue virus serotypes (DENV 1-4) that are transmitted primarily by the female *Aedes aegypti* mosquito, with *Aedes albopictus* as a secondary vector. Common disease manifestations range from asymptomatic to moderate febrile illness, with a smaller proportion of patients who progress to severe illness characterized by hemorrhage, shock and death [4]. Integrated vector control and surveillance remain the principle strategies for disease prevention and control in endemic regions, as no vaccine or specific medical treatment are yet available. Macro social and environmental drivers have facilitated the global spread and persistence of dengue, including growing vulnerable urban populations, global trade and travel, climate variability, and inadequate vector control [5-8]. However, we have a limited understanding of the relative effects of these drivers at the local level, restricting our ability to predict and respond to site-specific dengue outbreaks.

Early warning systems (EWS) for dengue and other climate-sensitive diseases are decision-support tools that are being developed to improve the ability of the public health sector to predict, prevent, and respond to local disease outbreaks [9,10]. An EWS incorporates environmental data (e.g., climate, altitude, sea surface temperature), epidemiological surveillance data, and other social-ecological data in a spatiotemporal prediction model that generates operational disease risk forecasts, such as seasonal risk maps. Previous studies have demonstrated the utility of this approach for vector-borne diseases, including for dengue [11-13], malaria [14-16] and rift valley fever [17]. Maps and other model outputs are linked to an epidemic alert and response systems, triggering a chain of preventive interventions when an alert threshold is reached.

One of the first steps in developing an EWS is to characterize the spatiotemporal dynamics and the covariates associated with historical disease transmission. This is often done by developing GIS base maps of epidemiological, environmental, and social data to identify risk factors; and through time series analyses of epidemiological and climate data. These analyses require cross-institutional integration of expertise and data, including epidemiological and entomological data from ministries of health, climate information from national institutes of meteorology, and social-ecological spatial data from national census bureaus. Previous studies indicate that associations among climate, socioeconomic indicators and dengue risk vary by location and time, indicating the need for analyses of dengue risk that consider the local context to explain transmission mechanisms [18-24]. Importantly, these analyses also need to consider the spatial and temporal scales of ongoing data collection and surveillance activities to ensure that the model outputs can support an operational EWS.

The National Institute of Meteorology and Hydrology (INAMHI) of Ecuador is coordinating efforts with the Ministry of Health (Ministerio de Salud Pública – MSP) to develop an operational dengue EWS for coastal regions of Ecuador, where the disease is hyper-endemic [25]. Our previous studies in southern coastal Ecuador demonstrated the potential to develop a dengue EWS that incorporates climate and non-climate information. We found that the magnitude and timing of dengue outbreaks were associated with anomalies in local climate, the El Niño Southern Oscillation (ENSO), the virus serotypes in circulation, and vector abundance [26]. Local field studies showed that dengue risk also depended on household risk factors (e.g., access to piped water infrastructure, demographics, water storage behaviors, housing conditions) [22]. Our recent advances in seasonal climate forecasts indicate that the forecasts in this region have considerable skill (i.e., predictive ability) [27,28]. Building on these previous studies, this study was conducted to characterize the spatiotemporal dynamics, climatic and social-ecological risk factors associated with the largest dengue epidemic (2010) on record in the coastal city of Machala, Ecuador, an important site for dengue surveillance in the region.

## Methods

### Study Area

Machala, El Oro Province, is a mid-sized coastal port city (pop. 241,606) [29] located in southern coastal Ecuador, 70 kilometers north of the Peruvian border and 186 kilometers south of the city of Guayaquil (the epicenter of historical dengue outbreaks in the region). Dengue is an emerging disease in this region, with the first cases of dengue hemorrhagic fever (DHF) reported in 2005. The disease is now hyper-endemic, with year round transmission and co-circulation of all four serotypes. Recent multi-country studies showed that Machala had the highest *Ae. aegypti* larval indices of ten sites in other countries in Latin America and Asia [30,31] Given the high burden of disease, the high volume of people and goods moving across the Ecuador-Peru border, and proximity to Guayaquil, Machala is a strategic location to monitor and understand dengue transmission dynamics.

In 2010, DENV-1 caused the largest dengue epidemic to date, with over 4,000 cases reported in El Oro province [32] (Figure 1A). In Machala, there were 2,019 cases of dengue fever (and 77 DHF) or an incidence of 84 dengue cases (and 3 DHF cases) per 10,000 population per year, compared to 25 dengue cases per 10,000 population per year from 2003 to 2009. The greatest burden of disease (58%) during the epidemic was among individuals under 20 years of age (Figure 2). The number of cases in older adults (i.e., 11% of cases reported by people over 50) indicated a strong force of infection, and that dengue was a relatively new disease in the population. The epidemic occurred during a wetter than average year, with elevated vector indices. Cumulative rainfall from January to April 2010 was 56% above the 1986 to 2009 average. The percent of households with *Ae. aegypti* juveniles (House Index) was 21.7 ± 4.11 (mean ± 95% CI) in 2010 compared to 14.3 ± 4.70 from 2003 to 2009 [33].

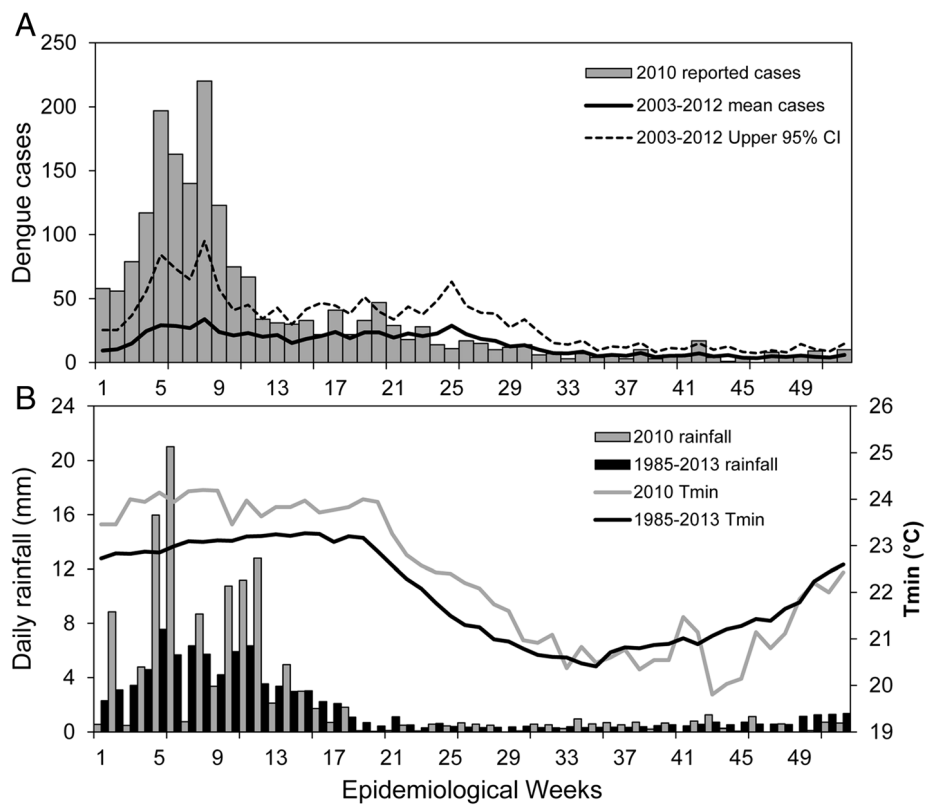

**Figure 1 Time series of dengue and local climate conditions in 2010 and historically in Machala, Ecuador. (A)** Weekly reported cases of dengue in 2010 and weekly average cases from 2003 to 2012; **(B)** weekly averages of rainfall and minimum air temperature (Tmin) in 2010 compared to the climatology (1986 to 2013 average conditions).

### Data sources

The following data from Machala were included in analyses: neighborhood-level georeferenced dengue cases, national census data, and entomological surveillance data from 2010; and time series of weekly dengue cases (aggregated to the city-level) and meteorological data from 2003 to 2012. These data were examined to identify potential social-ecological and climate variables associated with the presence of dengue fever during the 2010 outbreak in Machala, Ecuador. Epidemiological data were provided by INAMHI through a collaborative project with the MSP that was sponsored by the Ecuadorian government. Accordingly, no formal ethical review was required, as the data used in this analysis were de-identified and aggregated to the neighborhood- and city-level, as described below.

### Epidemiological data

INAMHI provided a map of georeferenced dengue cases from Machala in 2010, de-identified and aggregated to neighborhood-level polygons (n = 253) to protect the identity of individuals [34]. This map was generated from individual records of clinically suspected cases of dengue fever and DHF (aggregated as total dengue fever) reported to a mandatory MSP disease surveillance system, and the map included 83% of all dengue cases (n = 1,674) reported in 2010. Reported dengue cases were defined based on a clinical diagnosis. INAMHI also provided data for weekly dengue cases from Machala from 2003 to 2012 for the wavelet analysis described below.

### Social-ecological risk factors

We extracted individual and household-level data from the 2010 Ecuadorian National Census [29] to test the hypothesis that social-ecological variables were associated with the presence of dengue (Table 1). We calculated a composite normalized housing condition index (HCI) for each household by combining variables for the condition of the roof (CR), condition of the walls (CW), and condition of the floors (CF) (Equation 1). Each of the three variables ranged from 1 to 3, where 3 indicated poor condition. When summed, the values of the composite index ranged from 3 (min) to 9 (max), and we normalized the index from 0 to 1, where 1 indicated the worst housing condition.

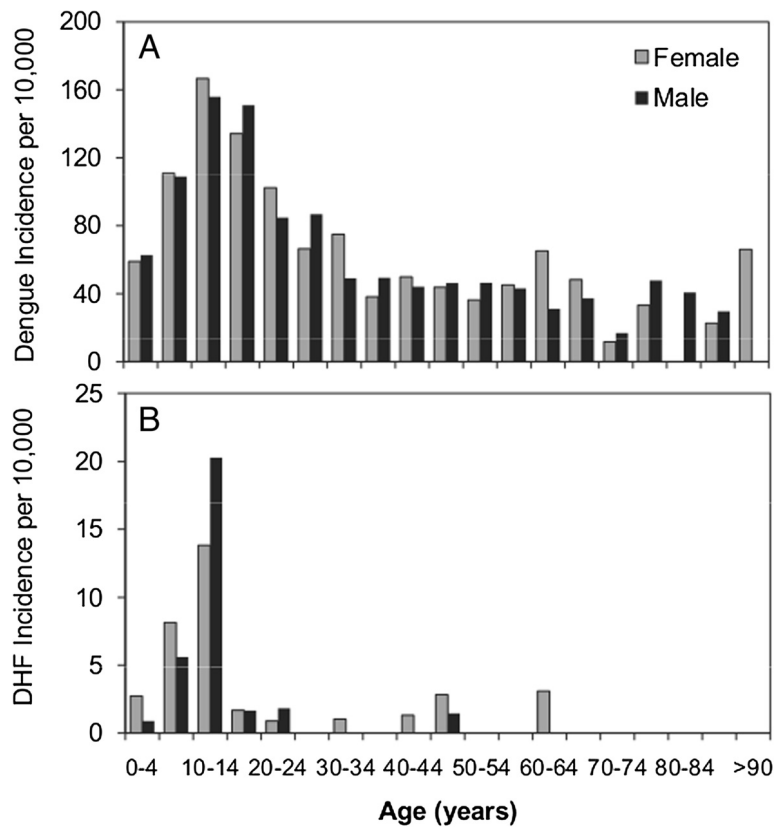

**Figure 2** Dengue incidence (per 10,000 population per year) by age and gender for (A) dengue fever and (B) dengue hemorrhagic fever in Machala in 2010.

$$HCI = [(CR + CW + CF) - \min] / (\max - \min) \quad (1)$$

Using individual and household census records, we recoded selected census variables and calculated parameters as the percent of households or percent of the population per census sector (n = 558 census sectors). The data element dictionary of recoded variables in Spanish is presented in Additional file 1: Table S1. To scale the sector-level polygon data to neighborhood-level polygons, we used the 'isectpolypoly' tool in Geospatial Modeling Environment [35,36]. We estimated neighborhood population by calculating the area-weighted sum, and estimated all other parameters by calculating area-weighted means. The neighborhood population estimates were also used to calculate neighborhood dengue prevalence and population density parameters.

### Entomological data
Vector surveillance data for *Ae. aegypti* from 2010 was obtained from the National Service for the Control of Vector-Borne Diseases of the MSP, and included quarterly House Indices (percent of households with *Ae. aegypti* juveniles) and Breteau Indices (number of containers with *Ae. aegypti* juveniles per 100 households). The average Breteau Index during the first two quarters of 2010 (January to June), was the vector index that was most strongly associated with dengue presence (1) or absence (0) (Pearson correlation, r = 0.2, $p$ = 0.001) and this period corresponded with the peak of the epidemic; accordingly, we selected this variable to test in the multivariate model (Table 1).

### Climate data
Daily meteorological data (rainfall and minimum air temperature) during the study period were provided by the Granja Santa Ines weather station located in Machala (3°17'16" S, 79°54'5" W, 5 meters above sea level) and operated by INAMHI. The weekly climatology (1985-2013) and weather during the study period are shown in Figure 1B. Weekly average rainfall and minimum temperature from 2003 to 2012 were included in the wavelet analysis, since it has been shown that these two climate variables explain an important part of the total variance of dengue cases in coastal Ecuador [26].

Table 1 Social-ecological parameters (mean and standard deviation - SD) tested in logistic regression models to predict dengue presence (1) and absence (0) at the neighborhood level in Machala in 2010

| Parameter | Mean | SD |
| --- | --- | --- |
| **Population density** | | |
| More than four people per bedroom (% households) | 14.6% | 6.4% |
| Population density (people per square kilometer) | 10,864 | 5,302 |
| More than one other household sharing the home (% households) | 2.2% | 1.6% |
| People per household | 3.88 | 0.52 |
| **Demographics** | | |
| Receive remittances (% households) | 10.8% | 3.2% |
| People emigrate for work (% households) | 2.2% | 1.2% |
| Mean age of the head of the household (years) | 45.2 | 3.0 |
| Head of the household has primary education or less (% households) | 35.9% | 12.9% |
| Afro-Ecuadorian (% population) | 9.6% | 6.9% |
| Head of the household is unemployed (% households) | 23.0% | 5.3% |
| Head of household is a woman (% households) | 30.3% | 4.5% |
| **Housing conditions** | | |
| Housing condition index (HCI), 0 to 1, where 1 is poor condition | 0.29 | 0.10 |
| No access to municipal garbage collection (% households) | 8.0% | 12.3% |
| No piped water inside the home (% households) | 34.4% | 18.7% |
| No access to sewerage (% households) | 22.4% | 27.6% |
| No access to paved roads (% households) | 26.7% | 22.3% |
| People drink tap water (% households) | 32.8% | 11.7% |
| Rental homes (% households) | 24.6% | 10.6% |
| **Other variables** | | |
| Average distance to the central hospital (km) | 2.36 | 1.27 |
| Average Breteau Index during the first two quarters of 2010 | 28.6 | 2.15 |

### Statistical analyses
#### Exploratory spatial analysis
We applied Moran's I with inverse distance weighting (ArcMap 10.1) to epidemiological dengue data from 2010 to test the hypothesis that dengue cases were randomly distributed in space. Moran's I is a global measure of spatial autocorrelation, that provides an index of dispersion from -1 to +1, where -1 is dispersed, 0 is random, and +1 is clustered. We identified the locations of significant dengue hot and cold spots using Anselin Local Moran's I (LISA) with inverse distance weighting (ArcMap 10.1). The LISA is a local measure of spatial autocorrelation [37] that identifies significant clusters (hot or cold spots) and outliers (e.g., nonrandom groups of neighborhoods with above or below the expected dengue prevalence). Previous studies have used Moran's I and LISA to test the spatial distribution of dengue transmission [38], including in Ecuador [39], allowing for comparison between studies.

#### Social-ecological risk factors
Census data aggregated to the neighborhood-level were examined to identify potential social-ecological variables associated with the presence of dengue fever, including population density, human demographic characteristics, and housing condition (Table 1). We hypothesized that the presence or absence of dengue was associated with one or more of these factors; each factor was presented as a suite of census variables, representing testable variable ensemble hypotheses in a model selection framework - a modeling strategy that has been previously described [22]. Variables for the average distance to the public hospital (Teofilo Davila Hospital, the provincial hospital located in the city center) and the Breteau

Index were also tested in the model to assess geographic differences, potential underreporting, and other factors (e.g., microclimate, vector control) not captured by the census variables.

We centered all variables and selected the best-fit models (GLM, family = binomial, link = logit) using glmulti, an R package for multimodel selection [40]. All possible unique models were tested and ranked based on Akaike's Information Criterion (AIC) modified for small sample sizes (AICc) (Equation 2). We compared the top ranked model to the global model, which included all proposed variables as model parameters.

$$AIC = 2k - \ln(L)$$
$$AICc = AIC + \frac{2k(k+1)}{n-k-1} \quad (2)$$

Where $k$ is the number of parameters in the model, $n$ is the sample size, and $L$ is the maximized likelihood function for the model.

Parameter estimates and 95% confidence intervals (CI) were calculated for variables in the top ranked model (Table 2 Model A). Variance inflation factors (VIF) were calculated to assess multi-colinearity and model dispersion. We found that inclusion of the parameters for housing condition and piped water together led to overdispersion and highly inflated variance in the best-fit model (Table 2 Model B). Based on the strong linear correlation between housing condition and access to piped water (Figure 3), we replaced these variables with the residuals of housing condition regressed on access to piped water, and re-ran glmulti to identify the best-fit model. The inclusion of this variable enabled testing for the effect of housing condition beyond that which was explained only by access to piped water.

### Wavelet analysis

To understand the time-frequency variability of dengue and climate during the 2010 epidemic, we conducted a wavelet analysis of a 10-year time series of weekly incident dengue cases (2003-2012), rainfall and minimum temperature (Figure 4A). Wavelet analyses are ideal for noisy, non-stationary data, such as dengue cases data, which demonstrate strong seasonality and interannual variability (yearly changes) [41,42]. These analyses identify significant temporal scales (i.e., defined here as periods whose associated wavelet power is statistically significant for at least two continuous years; Figure 4) over time for a given variable, such as 2-year cycles or annual seasonal cycles of dengue transmission. Cross wavelet and wavelet coherency allowed us to compare two time series, such as climate and dengue, and to identify synchronous periods or signals.

The pre-processing of the time series data for Machala followed a two-step methodology described elsewhere in detail [27,28,43]. First, we quality-controlled the time series using a standard R package [44] to identify outliers and inconsistent values (e.g., minimum temperatures > maximum temperatures, negative precipitation values or negative frequency of dengue cases). Outliers were defined as data points at least three standard deviations above or below the mean. To account for real outliers (e.g., not artifacts produced by human, instruments, or transmission errors), we compared suspicious values with data from nearby climate stations. Entries that we

**Table 2 The parameters included in the best-fit logistic regression models to predict the presence (1) or absence (0) of dengue in neighborhoods in Machala in 2010**

| Parameter | Estimate | 95% CI | SE | VIF | P value |
|---|---|---|---|---|---|
| Model A. | | | | | |
| Intercept | 0.75 | 0.46 – 1.05 | 0.15 | | < 0.001 |
| Head of household is a woman | 7.77 | 0.73 – 15.17 | 3.67 | 1.14 | 0.034 |
| Age of head of household | 0.10 | 0.00 – 0.21 | 0.05 | 1.06 | 0.051 |
| Residual of HCI regressed on households with no access to piped water inside the home | 9.04 | 3.98 – 14.37 | 2.64 | 1.05 | < 0.001 |
| Distance to central hospital | −0.0005 | −0.0007 – 0.0 | 0.0001 | 1.20 | < 0.001 |
| Model B. | | | | | |
| Intercept | −7.59 | −14.24 – −1.32 | 3.28 | | 0.021 |
| Head of household is a woman | 7.30 | 0.11 – 14.83 | 3.74 | 1.18 | 0.051 |
| Age of head of household | 0.14 | 0.002 – 0.26 | 0.07 | 1.60 | 0.052 |
| No piped water inside the home | −3.18 | −6.08 – −0.4 | 1.44 | 3.64 | 0.027 |
| HCI | 9.16 | 4.06 – 14.56 | 2.66 | 3.11 | 0.001 |
| Distance to central hospital | −0.001 | −0.001 – 0.0 | 0.0 | 1.31 | <0.001 |

(Model A.) The best fit model, which included the residual of the HCI regressed on the variable for no access to piped water inside the home. (Model B.) The same model is shown with separate parameters for the HCI and no access to piped water inside the home to indicate the direction of the effects of the parameters in the model. VIF values indicate a high degree of multicollinearity in model B compared to model A. High values of HCI indicate poor housing condition.

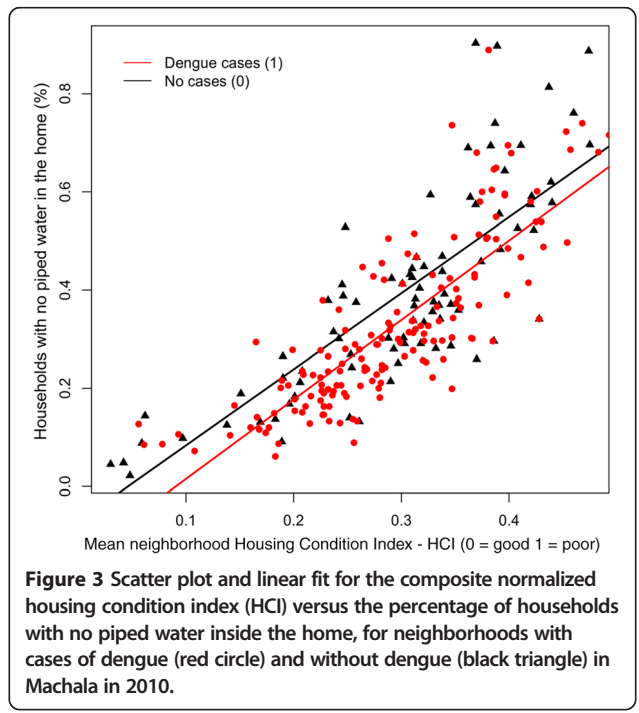

**Figure 3 Scatter plot and linear fit for the composite normalized housing condition index (HCI) versus the percentage of households with no piped water inside the home, for neighborhoods with cases of dengue (red circle) and without dengue (black triangle) in Machala in 2010.**

deemed to be uncorrectable were flagged as missing values. Then we used the R package 'RHTestsV4' [45-48] to detect and correct temporal inhomogeneities in these variables. The climate time series did not need substantial corrections. Weekly dengue case data were transformed to weekly incidence using a linear interpolation of local population data from the 2001 and 2010 national censuses. The final step in data pre-processing involved the normalization of the three variables to constrain variability. Dengue incidence and rainfall time series had non-normal probability density functions, thus they were percentile-transformed [34].

The wavelet analysis of dengue and climate data enabled us to identify common periodicity patterns (e.g., annual or biannual signals) and anomalous climate conditions during the 2010 outbreak. We used Morlet wavelets in Matlab [49] to compute the time series' wavelet power spectrum and to identify significant periods for each variable, cross-wavelet power to identify periods where dengue-rainfall and dengue-temperature had high common power, and the coherence spectra to identify local co-variability of dengue-rainfall and dengue-temperature [50]. Significance testing ($p \leq 0.05$) was conducted using an AR1 background noise for the first two spectra, and a Monte Carlo approach to compute the significance levels in the coherence spectrum. Statistically significant regions are displayed enclosed by a solid black line in the wavelet plots; and cones of influence (COI), where edge effects increase the uncertainty of the analysis, are shown as a lighter shaded region (Figures 4, 5 and 6). The arrows represent the relative phase, which is indicative of the lags between the two time series, as determined by frequency and time [51].The direction of the arrows can be used to quantify the phase relationship: arrows pointing horizontally to the right (left) indicate that the two variables are in (anti-) phase. When the signals of two time series are in phase, their maximum amplitudes occur simultaneously.

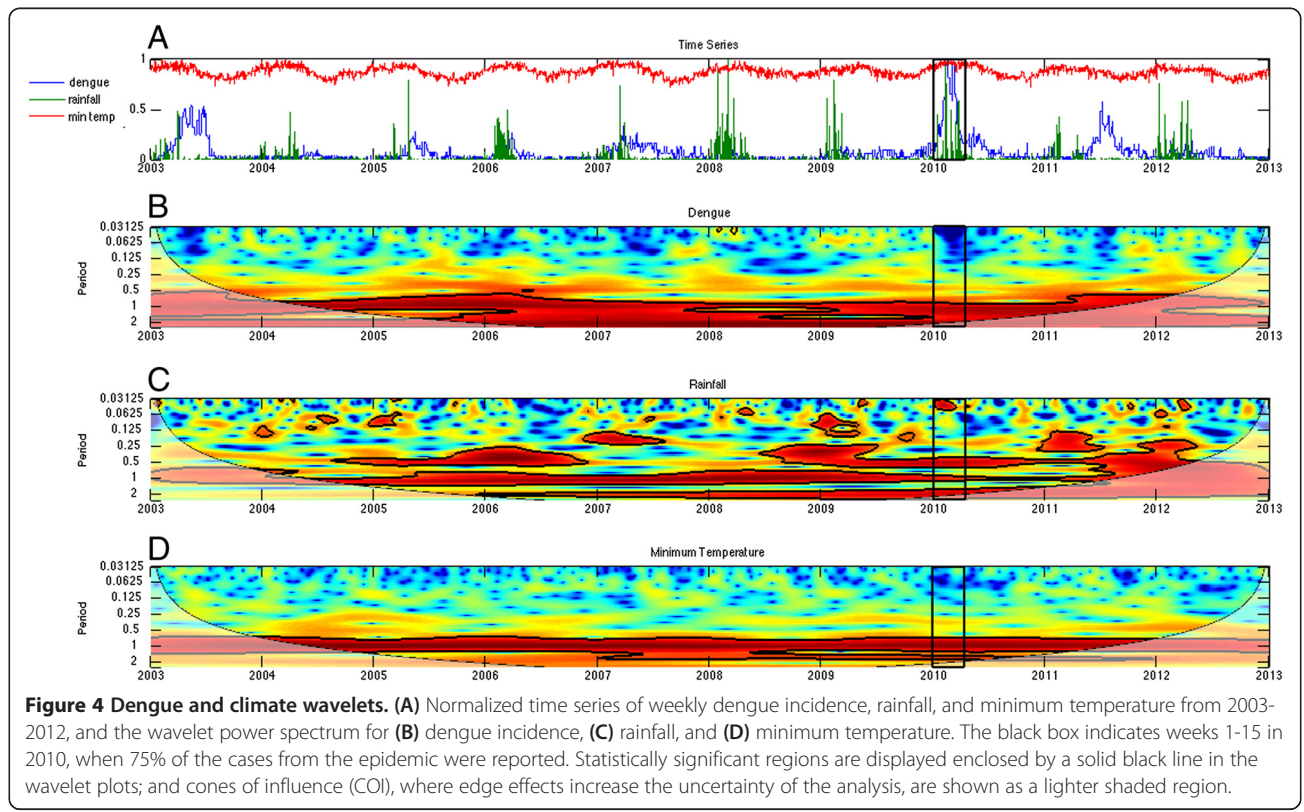

**Figure 4 Dengue and climate wavelets. (A)** Normalized time series of weekly dengue incidence, rainfall, and minimum temperature from 2003-2012, and the wavelet power spectrum for **(B)** dengue incidence, **(C)** rainfall, and **(D)** minimum temperature. The black box indicates weeks 1-15 in 2010, when 75% of the cases from the epidemic were reported. Statistically significant regions are displayed enclosed by a solid black line in the wavelet plots; and cones of influence (COI), where edge effects increase the uncertainty of the analysis, are shown as a lighter shaded region.

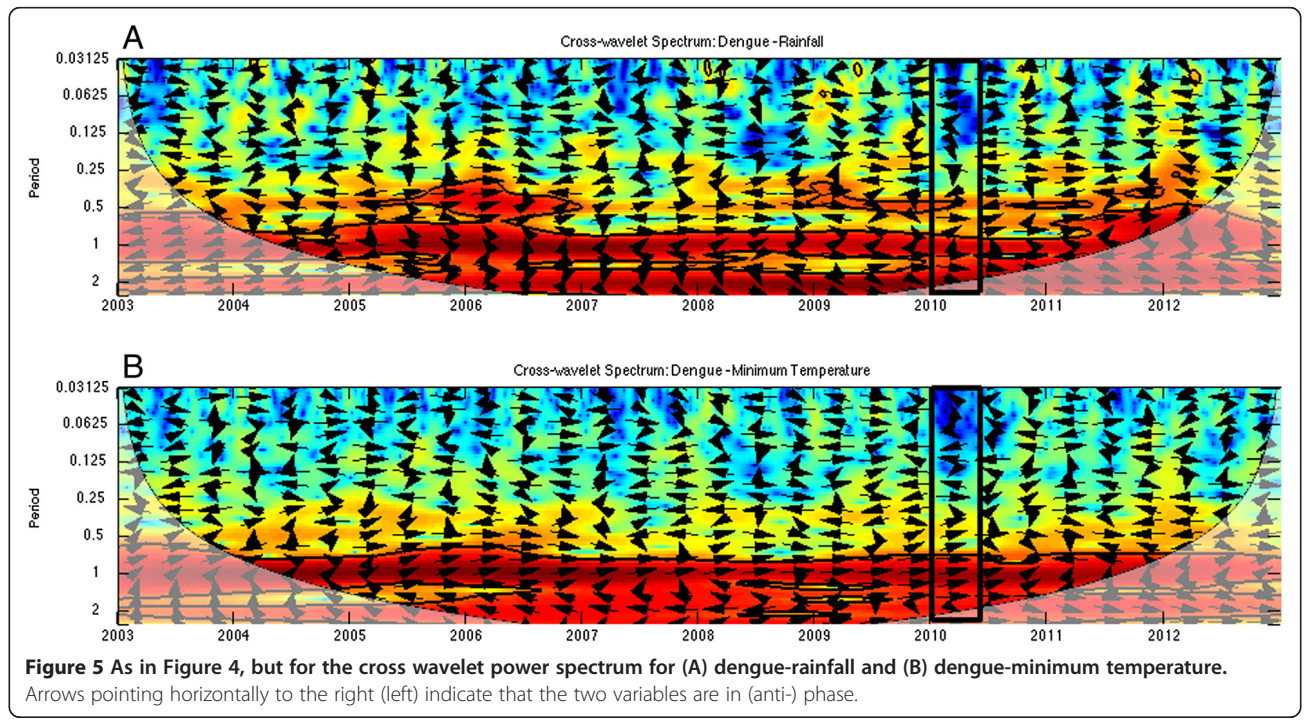

**Figure 5** As in Figure 4, but for the cross wavelet power spectrum for (A) dengue-rainfall and (B) dengue-minimum temperature. Arrows pointing horizontally to the right (left) indicate that the two variables are in (anti-) phase.

## Results

### Spatial analyses and social-ecological risk factors

Average neighborhood dengue incidence in 2010 was 76.7 ± 14.3 (95% CI) per 10,000 population (range: 0 to 775.8) (Figure 7A). The distribution was heavily left skewed, with 35% of neighborhoods (n = 89) reporting zero cases (Additional file 2: Figure S1). Dengue cases during the epidemic were significantly clustered (Moran's I = 0.03, $p$ <0.001). Findings from the LISA analysis indicated that there were significant dengue hotspots (n = 15 high-high neighborhoods) in west-central Machala, and a smaller number of significant outliers (n = 2 high-low neighborhoods, n = 5 low-high neighborhoods) ($p$ <0.05, Figure 7b).

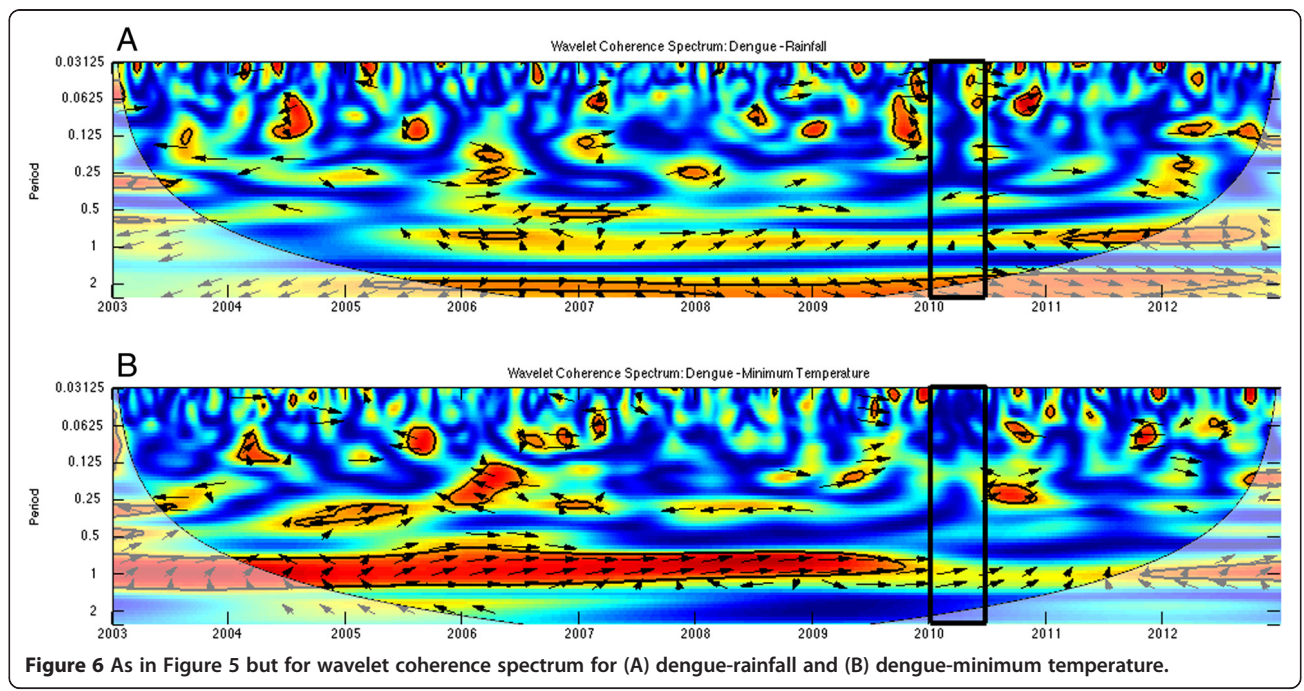

**Figure 6** As in Figure 5 but for wavelet coherence spectrum for (A) dengue-rainfall and (B) dengue-minimum temperature.

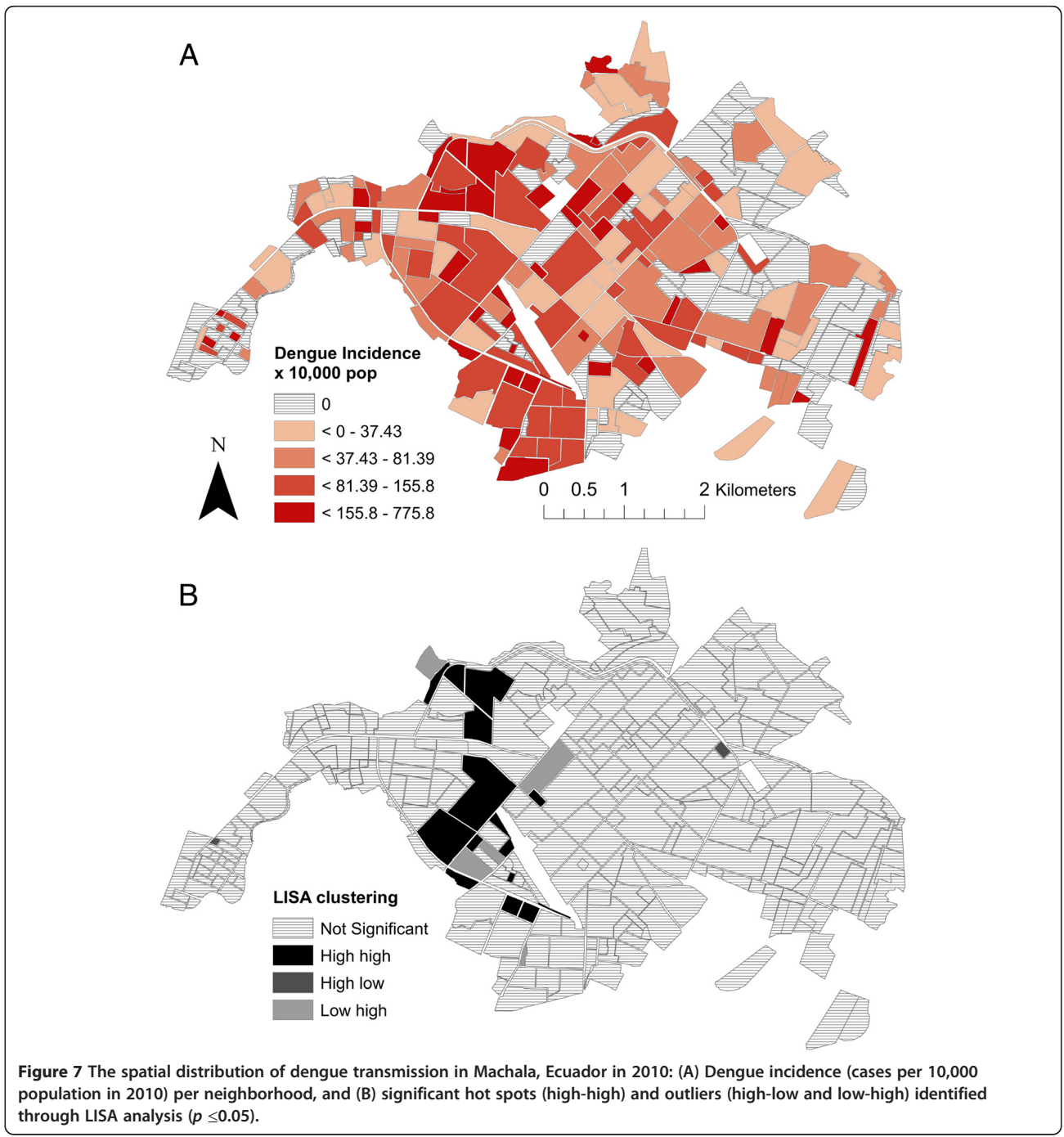

**Figure 7 The spatial distribution of dengue transmission in Machala, Ecuador in 2010:** (A) Dengue incidence (cases per 10,000 population in 2010) per neighborhood, and (B) significant hot spots (high-high) and outliers (high-low and low-high) identified through LISA analysis ($p \leq 0.05$).

The top ranked model to predict the presence of dengue, identified through the multimodel selection process, was a better fit than the global model that included all of the proposed social-ecological variables (global model AICc = 311, top ranked model AICc = 291.3, ΔAICc = 19.7). The top ranked model included the residual of the HCI regressed on no access to piped water, distance from the central hospital, and demographics of the heads of households (i.e., older age, female gender) (Table 2 Model A, Figure 8, Additional file 3: Figure S2). We also presented the model with the HCI and access to piped water as separate variables, to indicate the direction of the effect of each variable (Table 2 Model B). Neighborhoods were more likely to report dengue if they had poor housing condition and had greater access to piped water inside the home. When we compared neighborhoods with similar housing conditions, neighborhoods were more likely to report dengue cases if they had greater access to piped water inside the home (Figure 3). Inclusion of the residual variable in the model (Table 2 Model A) reduced multicollinearity, as indicated by the low VIFs.

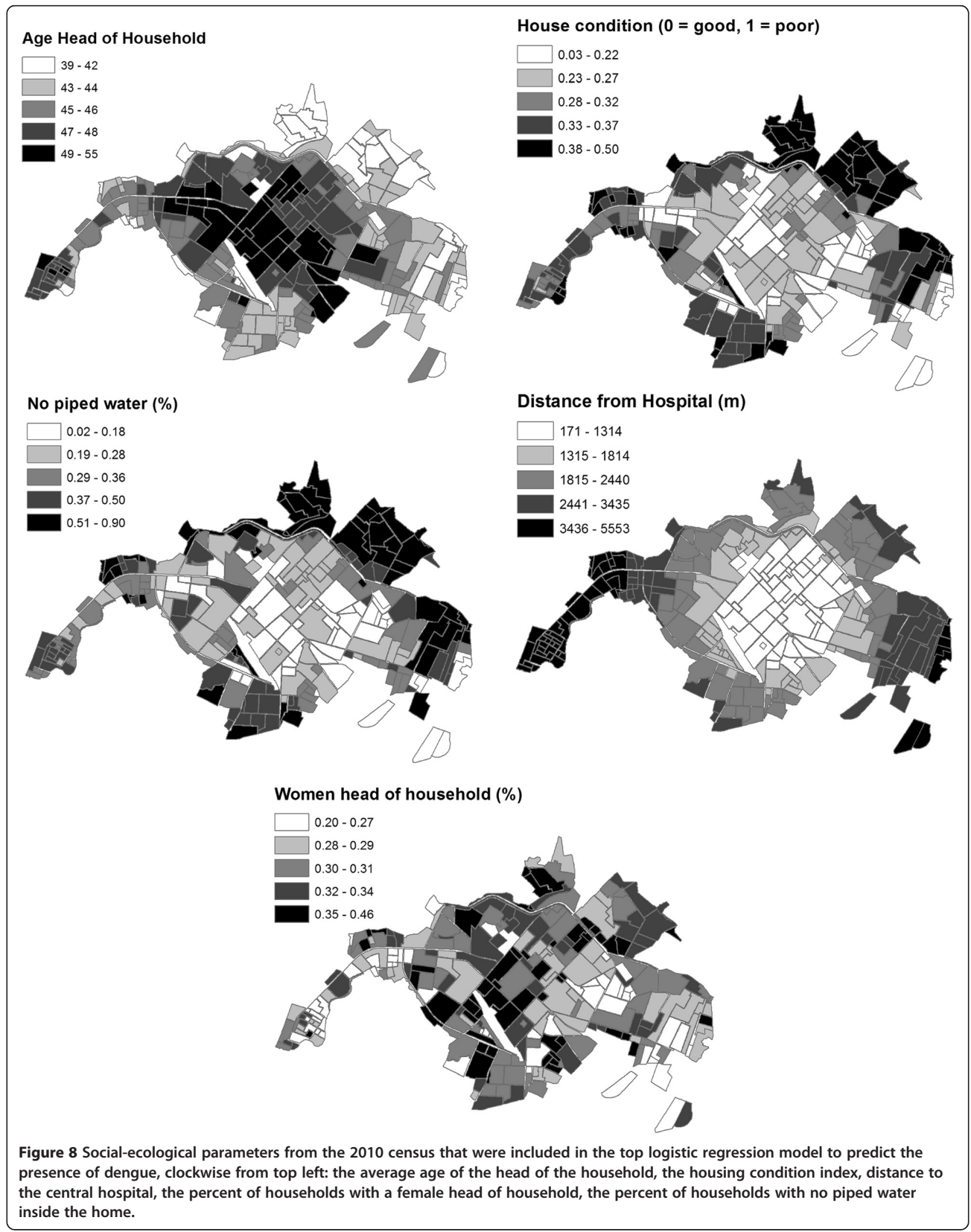

**Figure 8** Social-ecological parameters from the 2010 census that were included in the top logistic regression model to predict the presence of dengue, clockwise from top left: the average age of the head of the household, the housing condition index, distance to the central hospital, the percent of households with a female head of household, the percent of households with no piped water inside the home.

Multiple best-fit models were within the predetermined threshold criteria of ΔAICc ≤2 of the top model and weights greater than 1.5% (Additional file 4: Table S2). In addition to the parameters included in the top model, the following variables were included in competing best-fit models: Breteau Index, population density, households with people who emigrate for work, and households without access to paved roads.

### Temporal climate analyses

We found that multiple temporal scales were involved in local dengue transmission dynamics, as shown in the wavelet power spectrum for dengue incidence (Figure 4B). In wavelet analyses, strong significant signals at a certain frequency are associated with persistent (quasi) periodic cycles in the time series (e.g., a 1-year band indicates presence of annual cycles). There was a strong and significant signal for the ~2-year periodic band for dengue incidence. There was also a significant signal for the ~1-year periodic band, although it was less frequent (e.g., 2003, 2006, and 2011). Signals around and above the 4-year periodic band were not considered, as they fell inside of the COI (Figure 4B). These results suggest that dengue periodicity in this locality is not only annual (~1 year), but that there is also an important biannual cycle (~2 year), that may reflect typical time scales of extrinsic (e.g., climate) and intrinsic (e.g., immunologic) processes involved in the occurrence of dengue for this region.

The rainfall and minimum temperature spectra in Figure 4C,D demonstrated a strong annual signal (1-year periodic band), in agreement with the annual dengue cycle in the region. There was no evidence of relevant changes in variability for minimum temperature; the corresponding signal in the power spectrum is continuous around the annual band. In contrast, there were fluctuations in the ~1-year band for precipitation, likely associated with periods of low precipitation. Beginning around 2006, both climate variables demonstrated significant power around the 2-year band, a feature that is most noticeable in the rainfall data, particularly in recent years.

The cross-wavelet power spectra for dengue and rainfall (Figure 5A), and dengue and minimum temperature (Figure 5B) showed regions in the time-frequency space with high common power in the 1-year and 2-year bands, suggesting a relationship between climate and dengue incidence at both time scales. The corresponding wavelet coherence spectra, however, indicated that the dengue and rainfall co-vary mostly in the 2-year band (Figure 6A), while dengue and minimum temperature co-vary mostly in the 1-year band (Figure 6B). This suggests that temperature and rainfall have well-differentiated roles in dengue transmission. The directions of the arrows in the plots indicate a slow change of phase in the co-variability of dengue and rainfall in the 2-year band, approaching in-phase behavior in late 2009, and we observed synchronized co-variability for the 1-year band in 2009 and at the start of 2010. These results highlight the distinct roles of these climate variables in dengue transmission at different temporal scales, and the importance of the phase and timing of climate variables with respect to dengue transmission.

We found that the 2010 epidemic episode could be characterized by a combination of annual and bi-annual signals in dengue transmission and climate variables. The outbreak was characterized by a combination of in-phase variability of above normal minimum temperatures, and quasi-in-phase above normal rainfall episodes associated with the late 2009 to early 2010 moderate El Niño event (see arrows pointing right along the 1-year band at the bottom of Figures 5 and 6). The times series (1995-2010) of monthly anomalies in dengue cases from El Oro province, ENSO, temperature and rainfall have been previously described (See Figure 2 in Stewart Ibarra & Lowe 2013) [26]. This analysis demonstrated that the observed effect (quasi-simultaneity in the variability of dengue, temperature and rainfall) was present in early 2010, but not in any other year of the period under study (Figures 5 and 6).

### Discussion

Dengue is the most important mosquito-borne viral disease globally, and has increased in incidence and distribution despite ongoing vector control interventions during the last three decades [1-3]. To date, we have a limited understanding of the spatiotemporal dynamics of dengue transmission, particularly at the local scale, due to the complex, non-stationary relationships among dengue infection, climate, vector, and virus strain dynamics [41,52-54]; and the geographic and temporal variation in the social-ecological conditions that influence risk [18-21]. More robust analysis tools, such as wavelet analyses and multimodel inference, and the increasing availability of geospatial epidemiological, climate, and social-ecological data have increased our ability to explore these dynamics. Studies such as this provide critical information to improve disease surveillance and to develop an EWS and other evidence-based interventions.

In this study, we found that neighborhoods with certain social-ecological conditions were more likely to have cases of dengue during the largest outbreak to date in El Oro Province. Dengue cases were clustered in neighborhood-level transmission hotspots near the city center during the epidemic. Risk factors included poor housing condition, greater access to piped water inside the home, less distance to the central hospital, and demographics of the heads of households (*i.e.*, older age, female gender). In analyses of 10 years of weekly epidemiological and climate data, we found that dengue, rainfall and minimum temperature co-varied and had common power at 1-year and 2-year cycles, with quasi-synchronized higher than average rainfall and minimum temperatures likely contributing to the 2010 dengue outbreak. This study contributes to ongoing efforts by INAMHI and the MSP of Ecuador to develop a dengue prediction model and early warning system. Findings from this study will inform the development of dengue vulnerability maps and climate-driven dengue seasonal forecasts that provide the MSP with information to target high-risk regions and seasons, allowing for more efficient use of scarce resources [9].

Spatial dynamics and social-ecological risk factors
In Machala, a relatively small and heterogeneous city, there was evidence of unequal exposure or unequal reporting of dengue. During the epidemic, dengue transmission was focused in hotspots in the west-central urban sector, a middle- to low-income residential area with moderate access to urban infrastructure. Although people had access to basic services, our previous studies suggest that dengue control in these communities may be limited by the cost of household vector control, lack of social cohesion, and limited engagement with local institutions [55]. Previous studies that used spatial clustering statistics also found evidence of significant clustering of dengue transmission across the urban landscape [18,56-58]. A previous study in Guayaquil, Ecuador, identified neighborhood-level dengue hot and coldspots, and found that the location of hotspots shifted over the 5-year period, highlighting the spatially dynamic nature of dengue risk and the importance of multiyear studies [39]. Longitudinal field studies in Thailand found evidence of fine-scale spatial and temporal clustering of dengue virus serotypes and transmission at the school and household levels [59,60]. Focal transmission patterns are likely associated with the limited flight range of the *Ae. aegypti* mosquito. Recent studies in Peru demonstrated the importance of human movement patterns in determining spatial dengue transmission dynamics within an urban area [61,62]. At a regional scale, dengue outbreaks are likely influenced by human movement north and south along the Ecuador-Peru border. Future studies should continue to investigate the regional effects of cross-border movement of people and goods, and the local effects of intra-urban movement between work, school, and home to better understand the spatial dynamics of dengue transmission.

We found that the combination of HCI and access to piped water was the most important risk factor for dengue transmission, as indicated by the magnitude of the best-fit model parameter estimate (Table 2 Model A); this parameter was also a significant variable in all other top models (Additional file 4: Table S2). Neighborhoods were more likely to report dengue if they had poor housing conditions (likely associated with lower income) and greater access to piped water inside the home (likely associated with older, established communities with access to urban infrastructure). This apparent paradoxical relationship suggests that household water storage behaviors played an important role in the 2010 dengue outbreak. In our experience low-income households in Machala with access to piped water tend to store water in containers in the patio as a secondary water source, since water supply interruptions are common. These secondary water containers are often uncovered, and the containers become ideal *Ae. aegypti* larval habitat during the rainy season. In contrast, low-income households without access to piped water are likely to store water in containers as their primary water source (e.g., 55 gallon drums), frequently filling and emptying the containers and thus preventing *Ae. aegypti* from developing into adult mosquitoes.

Neighborhoods were also at greater risk of dengue if they were closer to the central hospital, reflecting either spatially biased reporting and/or a true increase in transmission near the city center. This variable was also significant in all of the top models (Additional file 4: Table S2). Given the small size of the city of Machala (~5 km across) and easy access to low-cost public transportation, travel time to the hospital was not likely to be a limiting factor. However, people from lower income communities may be less likely to seek medical care due to the cost of medicine and the high cost of missing work, leading to underreporting from the urban periphery. It is also possible that people residing near the city center in Machala were at greater risk because they may have been less willing to cooperate with vector control technicians (E. Beltran, *pers. comm.*), due in part to the misconception that dengue is a problem of poor communities at the urban periphery [55]. Households in these areas may also be at greater risk because they store water as a secondary water source, as described above. These findings highlight the complexity of the cultural and behavioral factors influencing dengue risk and the importance of local-level studies that consider the social context.

Our findings are consistent with a previous longitudinal field study of household risk factors for *Ae. aegypti* in Machala, where it was found that poor housing condition and access to piped water inside the home were positively associated with the presence of *Ae. aegypti* pupae [22]. This prior study found that *Ae. aegypti* were more abundant in the central urban area that had better access to infrastructure than in the urban periphery [22]. Interestingly, the same risk factors emerged in the study presented here and the prior field study despite differences in rainfall (i.e., the field study was conducted one year after the epidemic, during a drier than average year) and differences in spatial scale (i.e., household- versus neighborhood level). These findings indicate that high-risk households could be identified and targeted using a combination of census data and a locally adapted rapid survey of housing conditions, similar to the Premise Condition Index, an aggregate index measuring house condition, patio condition, and patio shade, which has been validated in other countries [63,64]. The HCI and the combined HCI-water access variables developed in this study should be explored and validated as dengue predictors in future studies in this region.

*Ae. aegpyti* juvenile indices were included in two of the top seven best-fit models to predict the presence of dengue in neighborhoods (Additional file 4: Table S2). A previous study in El Oro Province found that *Ae. aegypti* indices (House Index) were positively associated with dengue outbreaks at the province level [26]. Although pupal or adult indices are considered better predictors of dengue risk than larval indices [65], our findings suggest that larval indices may have some predictive power in this region. In Ecuador, entomological surveillance is limited to larval indices, and neighborhoods are rarely sampled in consecutive periods in a given year due to limited resources. These findings highlight the need for additional studies of the vector-dengue dynamics in this region and local evaluations of the robustness of vector abundance measures in order to strengthen cost-effective entomological surveillance systems.

### Climate and dengue periodicity

The wavelet analysis in this study provided a nuanced understanding of the relationships among local dengue transmission and climate variables at multiple temporal scales. The analysis of 10 years of weekly epidemiological and climate data from Machala provided evidence of significant 1-year and 2-year cycles in dengue, rainfall and minimum temperature. The 1-year cycles of minimum temperature and rainfall likely contributed to the annual dengue cycles observed in the power spectrum. This finding was expected, as previous studies have documented significant annual dengue cycles in this region [26]. Interestingly, we also found evidence of 2-year cycles in the rainfall wavelet power spectrum that were likely associated with biannual cycles of dengue transmission, a pattern that was previously undocumented in Ecuador.

Indeed, our analyses suggest that the 2010 dengue epidemic could be related to a timely coincidence of above normal minimum temperatures and above normal rainfall episodes during the moderate 2009 to 2010 El Niño event. Previous studies in this region have shown that *Ae. aegypti* abundance is associated with rainfall and minimum temperature [22]. In 2010, rainfall from February and March, the peak of dengue season, was almost double the long-term average (89% and 81% above average, respectively), likely increasing the availability of mosquito larval habitat. Temperature and temperature fluctuations influence rates of mosquito development and virus replication [66-71]. The slow rate of climate phase change observed in this analysis suggests the potential to monitor the climate in this region to identify future time periods with synchronous climate conditions similar to 2010, that may increase the risk of a dengue outbreak.

Our results indicate that the 2-year band in precipitation is an important component in the co-variability of dengue incidence for the period under study, although its role in the 2010 dengue epidemic requires further investigation. This periodic band is not unique to Machala. Two to three-year cycles of dengue transmission have been reported in other parts of the world [41,53], particularly in years associated with El Niño events. The statistically significant 2-year band is present in the dengue power spectrum for the entire time series. This is not true for rainfall or minimum temperature, whose variability in the region is strongly associated with El Niño-Southern Oscillation (ENSO). This suggests that although ENSO has a strong influence in the occurrence of dengue epidemics in coastal Ecuador, other variables (e.g., immunity) are also involved in the process and/or that there is a persistent mechanism for the climate's biannual contribution in the dengue spectrum. It is interesting to note that similar 2-year cycles have been reported for dengue and malaria in mountain locations in Peru [72], but not along the Peruvian Coast or Amazon [73]. We hypothesize that the biannual signal found in Peru and Machala is related to an additional climate mode present over the Andes in this region [28] in addition to ENSO. Machala may be uniquely situated to capture climate signals from ENSO and the so-called [28] Andean mode, given its proximity to the Andean foothills and the strong coupled climate-ocean system (i.e., teleconnections) present in the region.

### Limitations

Although this study revealed patterns of climate and social-ecological conditions as important drivers of dengue transmission, this study has some limitations. It should be noted that non-climate factors that were undocumented in this study (e.g., population immunity, vector control interventions) are also key drivers of interannual variability in dengue [26,74,75] and most likely influenced the 2010 outbreak. The 1-year of spatially explicit epidemiological data constrained our ability to assess whether the social-ecological factors associated with the spatial distribution of dengue transmission were consistent in time. The 10-year time series of weekly dengue data was not available at the appropriate spatial scale for this analysis. With multiple years of data, we could evaluate whether dengue transmission at the beginning of the dengue season or at the beginning of an epidemic is more likely to begin in neighborhoods with similar characteristics, to assess whether there are persistent high-risk, hotspot neighborhoods that trigger outbreaks. The analyses were also limited by a lack of laboratory confirmation for cases or information about the immune, nutritional, or health status of the population. We are currently collaborating with the MSP to improve dengue diagnostic infrastructure in the region and to reduce the time lag between epidemiological reporting and vector control interventions. Importantly, the MSP is undergoing a reorganization and decentralization process to merge the health and vector control divisions at the local level, with the goal of improving information flows and linking responses to evidence-based interventions.

## Conclusions

The results of this study highlight the importance of incorporating climate and social-ecological information with georeferenced and clinically validated epidemiological data in a dengue surveillance system. Investigators in Ecuador are exploring the development of web-based GIS for national dengue surveillance using open-access software. GIS is an effective tool to integrate diverse data streams, such as dynamic, real-time epidemiological and climate data with static vulnerability maps generated from census data. Open access tools are especially important in resource-limited settings, and analysis packages targeted to dengue are becoming available [76]. Web-based GIS tools have been developed for global dengue surveillance, such as the CDC's DengueMap, and for local dengue surveillance research projects [77,78]. National-level dengue GIS initiatives have been developed in countries such as Mexico [79], where Ministry of Health practitioners and software developers jointly designed the software platform. This collaborative approach to integrate diverse data streams will ideally provide public health decision-makers with information to assess intervention programs, allocate resources more efficiently, and provide the foundation for an operational dengue EWS.

## Additional files

**Additional file 1: Table S1.** Spanish dictionary of census variables evaluated in the multivariate model to predict the presence of dengue.

**Additional file 2: Figure S1.** Histogram showing the density distribution of neighborhood dengue incidence in Machala, 2010 (n = 253).

**Additional file 3: Figure S2.** Scatter matrix of parameters included in the top logistic regression model to predict the presence of dengue in neighborhoods in Machala in 2010.

**Additional file 4: Table S2.** Top competing logistic regression models (ΔAICc < 2 or Weight > 1.5%) from multi-model selection to predict the presence (1) and absence (0) of dengue at the neighborhood level in Machala in 2010.

#### Competing interests
The authors declare that they have no competing interests

#### Authors' contributions
AMSI, AGM, MJBC, and RM conceived of the investigation. EBA, TO, GCRC and KR compiled the data used in analyses. AMSI, AGM, and SJR conducted analyses and drafted the manuscript. All co-authors, AMSI, AGM, MJBC, JLF, RM, TO, GCRC, KR, assisted with interpretation of the data, provided feedback for this manuscript, and read and approved the final manuscript.

#### Acknowledgements
Many thanks to colleagues at the MSP and INAMHI for supporting ongoing climate–health initiatives in Ecuador. This work was funded by the National Secretary of Higher Education, Science, Technology and Innovation of Ecuador (SENESCYT), grant to INAMHI for the project "Surveillance and climate modeling to predict dengue in urban centers (Guayaquil, Huaquillas, Portovelo, Machala)," and the Global Emerging Infections Surveillance and Response System (GEIS), grant #P0001_14_UN. AGM used computational resources from the Latin American Observatory of Extreme Events (www.ole2.org) and Centro de Modelado Científico (CMC), Universidad del Zulia. The following institutes that participated in this study also form part of the Latin American Observatory partnership (http://ole2.org): International Research Institute for Climate and Society (IRI), Earth Institute, Columbia University, New York, NY, USA; and Centro de Modelado Científico (CMC), Universidad del Zulia, Maracaibo, Venezuela; Escuela Superior Politécnica del Litoral, Guayaquil, Ecuador; National Institute of Meteorology and Hydrology, Guayaquil, Ecuador.

#### Author details
[1]Department of Microbiology and Immunology, Center for Global Health and Translational Science, State University of New York Upstate Medical University, 750 East Adams St, Syracuse, NY 13210, USA. [2]International Research Institute for Climate and Society (IRI), Earth Institute, Columbia University, New York, NY, USA. [3]Centro de Modelado Científico (CMC), Universidad del Zulia, Maracaibo, Venezuela. [4]Department of Geography, Emerging Pathogens Institute, University of Florida, Gainesville, FL, USA. [5]School of Life Sciences, College of Agriculture, Engineering, and Science, University of KwaZulu-Natal, Durban, South Africa. [6]The National Service for the Control of Vector-Borne Diseases, Ministry of Health, Machala, El Oro Province, Ecuador. [7]Facultad de Medicina, Universidad Técnica de Machala, Machala, El Oro Province, Ecuador. [8]Escuela Superior Politécnica del Litoral, Guayaquil, Ecuador. [9]Division of Nutritional Sciences, Cornell University, Ithaca, NY, USA. [10]Center for Geographic Analysis, Harvard University, Cambridge, MA, USA. [11]National Institute of Meteorology and Hydrology, Guayaquil, Ecuador.

| Table S1. Spanish dictionary of census variables evaluated in the multivariate model to predict the presence of dengue. | | | | |
|---|---|---|---|---|
| Censo* | Variable original | Recodificacion | Filtro | Descripcion |
| P | P03 Cuantos años cumplidos tiene | n/a | | Promedio del edad de la casa |
| P | P03 Cuantos años cumplidos tiene | n/a | P02 = 1 | Promedio del edad del jefe de la casa |
| P | | | | Personas por hogar |
| P | P23 Cuales el nivel de instrucción más alto al que asiste o asistió | P23 <= 4 | P02 = 1 | Porcentaje de hogares donde el jefe del hogar tiene educación primaria o mes |
| P | P16 Como se identifica según su cultura y costumbres | P16==2 \| P16==3 \| P16==4 | | Porcentaje de la población que son Afro-Ecuatoriano, mulato o Negro |
| P | P27 Que hizo la semana pasado | P27 == 7 | P02 = 1 | Porcentaje de hogares con el jefe del hogar desempleado |
| P | P01 Cual es el sexo | P01 == 2 | P02 = 1 | Porcentaje de hogares con una mujer como jefa del hogar |
| V | V02 (estado del techo), V04 (estado de las paredes), V06 (estado del piso) | (sum (V02, V04,V06)-3)/6 | TOTPER>0 | Índice de condición del hogar normalizado (compuesto de 3 parametros, 0 -1, donde 1 = peor) |
| V | TOTPER Total de personas de la vivienda | TOTPER==0 | | Porcentaje de hogares desocupados |
| V | PERDOR Numero de personas pordormitorio | PERDOR >= 4 | | Porcentaje de hogares con más de 4 personas por dormitorio |
| V | V13 Principalmente como elimina la basura | V13 > 1 | TOTPER>0 | Porcentaje de hogares que eliminan la basura por carro recolector |
| V | V08 El agua que recibe la vivienda es: | V08 > 1 | TOTPER>0 | Porcentaje de hogares SIN agua por tubería dentro de la casa |
| V | V09 El servicio higiénico o escusado de la vivienda es | V09 > 1 | TOTPER>0 | Porcentaje de hogares SIN conexión al red público de alcantarillado |
| V | Vap: vía de acceso principal a la vivienda | VAP > 2 | TOTPER>0 | Porcentaje de hogares que tienen calle o carretera adoquinada, pavimentada o de concreto |
| V | V16 Cuantos grupos de personas(hogares) duermen en su vivienda y cocinan los alimentos por separado incluya su hogar | V16 == 2 | TOTPER>0 | Porcentaje de hogares que comparten su casa con más de otro hogar |
| V | VIVREM viviendas con remesas | VIVREM == 1 | TOTPER>0 | Porcentaje de hogares que reciben remesas |
| V | TOTEMI Total de migrantes | TOTEMI > 1 | TOTPER>0 | Porcentaje de hogares que tienen emigrantes |
| H | H06 Principalmente, el agua que toman los miembros del hogar | H06 < 5 | | Porcentaje de hogares que toman agua de la llave |
| H | H15 La vivienda que ocupa este hogares… | H15==6 | | Porcentaje de hogares que alquilan |

*P = censo de población, H = censo de hogar, V = censo de vivienda

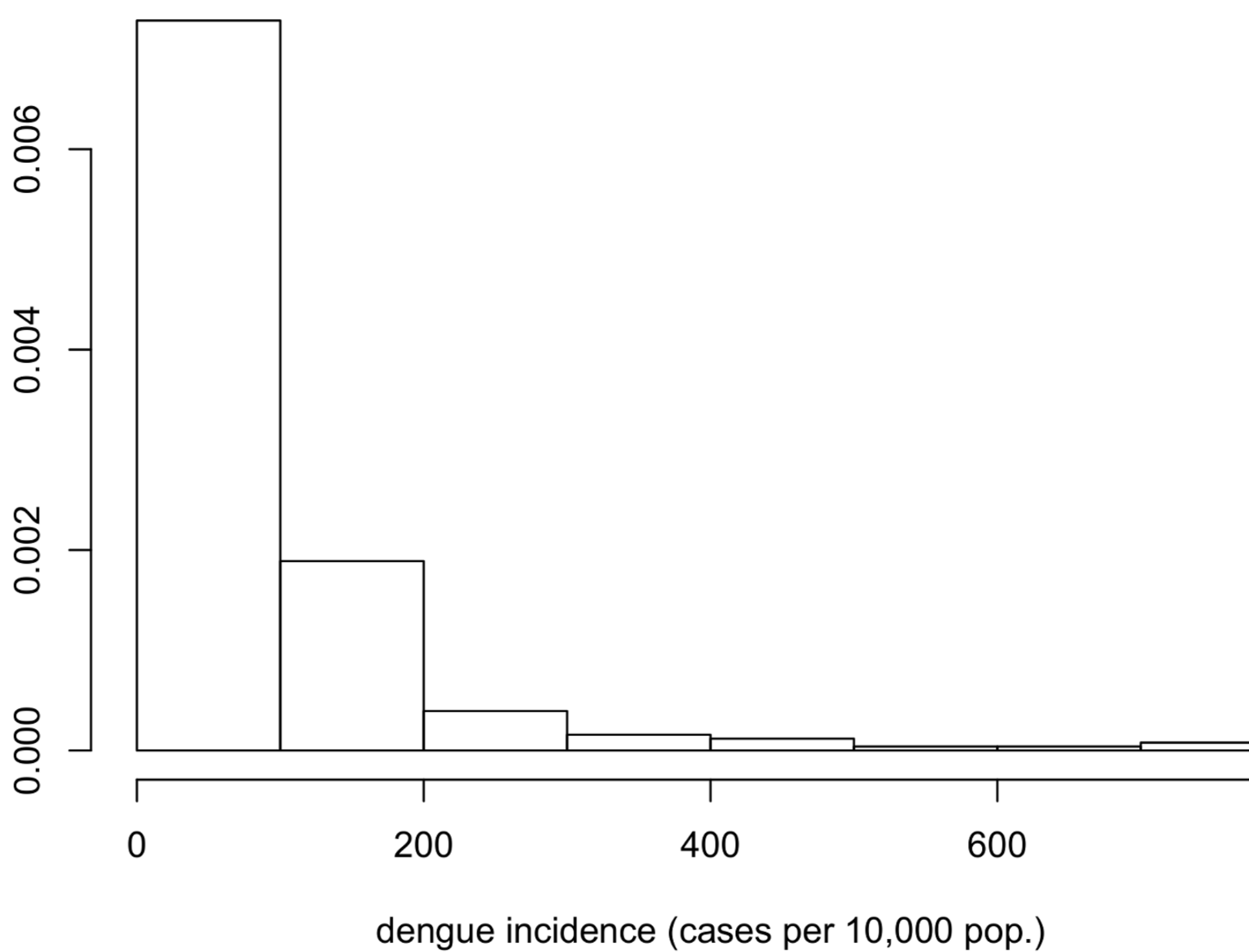

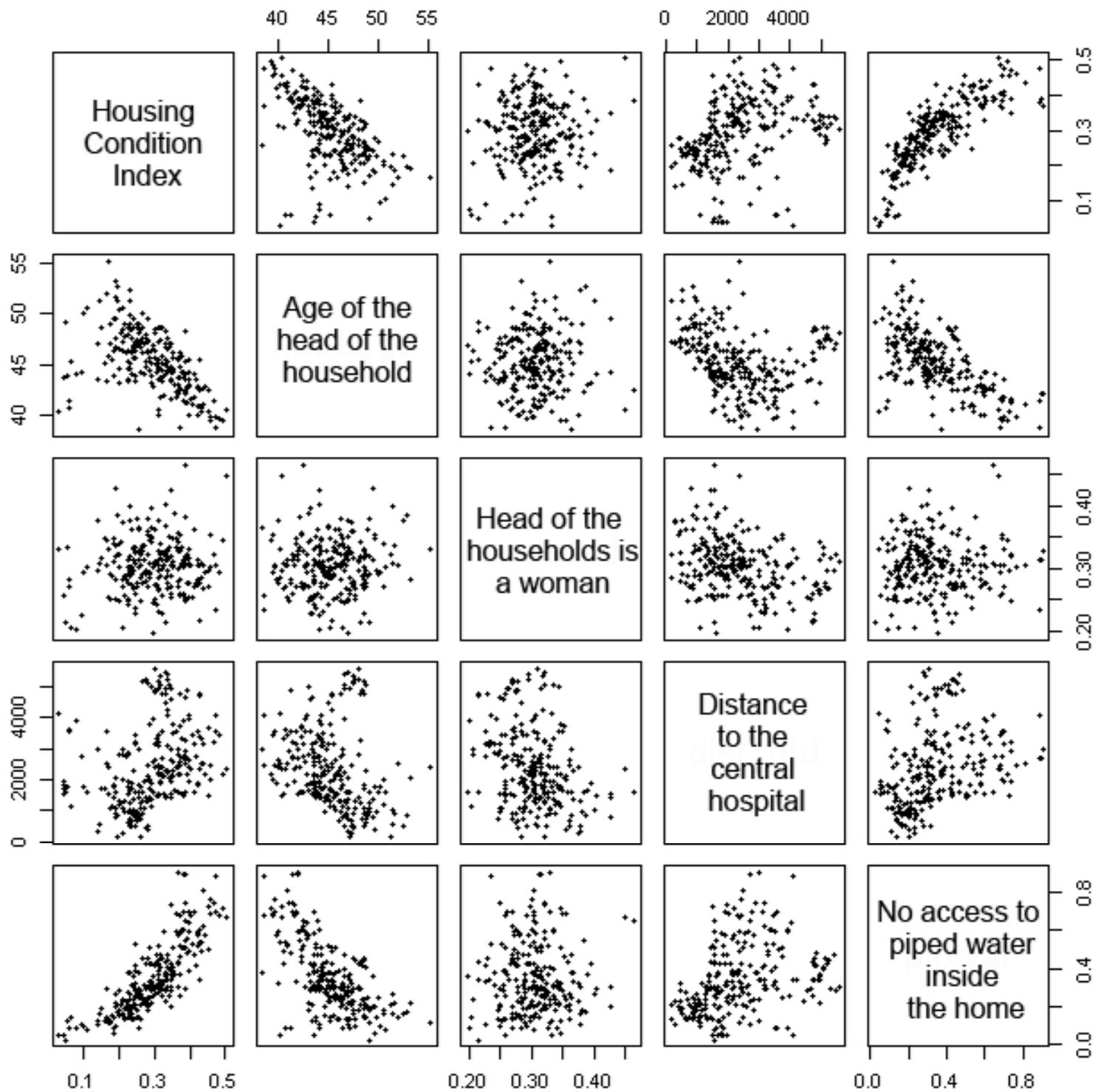

**Table S2. Top competing logistic regression models (ΔAICc < 2 or Weight > 1.5%) from multi-model selection to predict the presence (0) and absence (0) of dengue at the neighborhood level in Machala in 2010.***

| # | Model | AICc | Weights | ΔAICc |
|---|---|---|---|---|
| 1 | x ~ 1 + **wmhead** + headhh_age + **HCI_pipwat** + **dist_hosp** | 291.28 | 0.029 | 0.00 |
| 2 | x ~ 1 + **wmhead** + **headhh_age** + **HCI_pipwat** + **dist_hosp** + bi_12 | 291.87 | 0.020 | 0.59 |
| 3 | x ~ 1 + popdens + wmhead + headhh_age + **HCI_pipwat** + **dist_hosp** + bi_12 | 292.32 | 0.019 | 1.04 |
| 4 | x ~ 1 + popdens + wmhead + headhh_age + **HCI_pipwat** + **dist_hosp** | 292.22 | 0.018 | 0.95 |
| 5 | x ~ 1 + emigrt + wmhead + **HCI_pipwat** + **dist_hosp** | 292.39 | 0.017 | 1.12 |
| 6 | x ~ 1 + **wmhead** + **headhh_age** + pave + **HCI_pipwat** + **dist_hosp** | 292.62 | 0.017 | 1.35 |
| 7 | x ~ 1 + emigrt + wmhead + headhh_age + **HCI_pipwat** + **dist_hosp** | 292.49 | 0.016 | 1.21 |

*Significant parameters ($P \leq 0.05$) are bolded.

**Parameter dictionary**

| | |
|---|---|
| wmhead | = Head of household is a woman (% households) |
| headhh_age | = Mean age of the head of the household |
| HCI_pipwat | = Housing condition index (HCI) regressed on households with no access to piped water insides the home |
| dist_hosp | = Average distance to the central hospital (km) |
| bi_12 | = Average Breteau Index during the first two quarters of 2010 |
| popdens | = Population density (people per square kilometer) |
| emigrt | = People emigrate for work (% households) |
| pave | = No access to paved roads (% households) |